# Mathematical model of the Lamé Problem for Simplified Elastic Theory applied to Controlled-Clearance Pressure Balances

Saragosa Silvano[1)]

[1)]Department of Education and Science, Italy

**Abstract.** In order to obtain more accurate measurements, the use of pressure balances with the controlled-clearance system is well known in the history of high pressures metrology. Furthermore, the introduction of assemblies with cylinders made up of two materials to the construction of controlled-clearance pressure balances, is one of the most important developments in the field of high pressure measurements.
The use of some analytical procedures for the characterization of the pressure balances (also known as piston gauges) becomes more difficult especially for the case of geometries with composite cylinders. Those geometries were also studied with the analytical method developed by the author[1] (based on a simplified model) and described in this paper. This method was also reported in some not original publications from 2003 to 2007.
The analysis begins with the mathematical model of thick-walled cylinder applied to the Mechanical theory of elastic equilibrium for the formulation of the so called *Simplified Elastic Theory* which represents an analytical approach for the study of pressure balances. This analysis is known as the *Lamé problem*. The solution of the Lamé Problem is used to determine the pressure distortion coefficient $\lambda$ for the controlled-clearance pressure balances including the case of pressure balances with the cylinder made up of two materials.

## 1. Introduction

One of the most important problems in order to improve the performances of the pressure balances (especially in the field of high pressure measurements) is the accurate calculation of the pressure distortion coefficient $\lambda$. For this purpose, the methods can be classified as theoretical or experimental[6]. Furthermore, according to Dadson *et al.*, the effect of the measured pressure $P$ on the effective area of a piston-cylinder assembly $A_e$ is described by the well known relationship:

$$A_e = A_0(1 + \lambda P)$$

Where $A_0$ is the zero-pressure effective area.

A theoretical method described in this paper is *the simplified elastic theory*. It is well known that Johnson, Newall and Tsiklis (1953-68) have used some relationships based on the simplified elastic theory for the analytical evaluation of the effective area $A_e$ of the piston-cylinder units (PCUs). But those equations are valid in the case of pressure balances with cylinder made up of single material. A summary of this work is reported in[6] for both the simple and re-entrant units.

Under the same hypotheses used for the simple and reentrant units, we can determine the pressure distortion coefficients of controlled-clearance (CC) PCUs including also the case of composite cylinder [1].

The analytical method described in this paper was reported in the following publications:

- [2] is the first publication based on the thesis [1]: this work reports (extracted from [1]) the analytical demonstrations and the two formulae obtained by equations (23) and (28) for the evaluation of the pressure distortion coefficient of controlled-clearance piston gauges (CCPGs) with the cylinder made up of both single material and two materials respectively; this theoretical aspect was also reported in [10].

- In [3] there are the two equations mentioned above: in chapter 5.3 (from page 94 to page 100) entitled "The simplified method based on Lamé equation", the two formulae (eqs. (23) and (28) given in [1]) are reported to evaluate the pressure distortion coefficient of CCPGs.

The analysis of this paper is based on the following steps:

I) differential formulation of the *Lamé Problem*: Lamé theory of the thick-walled cylinder applied to the Mechanical theory of elastic equilibrium (section 2);

II) choice of the model: the model is important to schematize the piston-cylinder assembly; a wrong model (as reported in [2,3]) leads to erroneous formulae for $\lambda$ (sections 3 and 4);

III) boundary conditions are necessary to solve the *Lamé Problem*; the solution of the *The Lamé Problem* allows us to determine the radial displacements for analysing the particular piston-cylinder assembly (sections 3.1 and 4.1);

IV) on the basis of the hypotheses derived from [6] we can evaluate the pressure distortion coefficient for the CCPGs also for the case of cylinder made up of two materials [1](section 4.2).

## 2. The Lamé Problem

Among the theoretical methods, the Simplified Elastic Theory is an analytical approach to evaluate the pressure distortion coefficient by the solution of a system of differential equations called the *Lamé equations*. The results of this analysis allow one to the determination of the stresses and distortions for the case of thick walled cylinder. The method is formally known as *The Lamé Problem* and the differential formulation of this problem is based on the analysis of the equilibrium of an elementary volume [5] (fig.1).

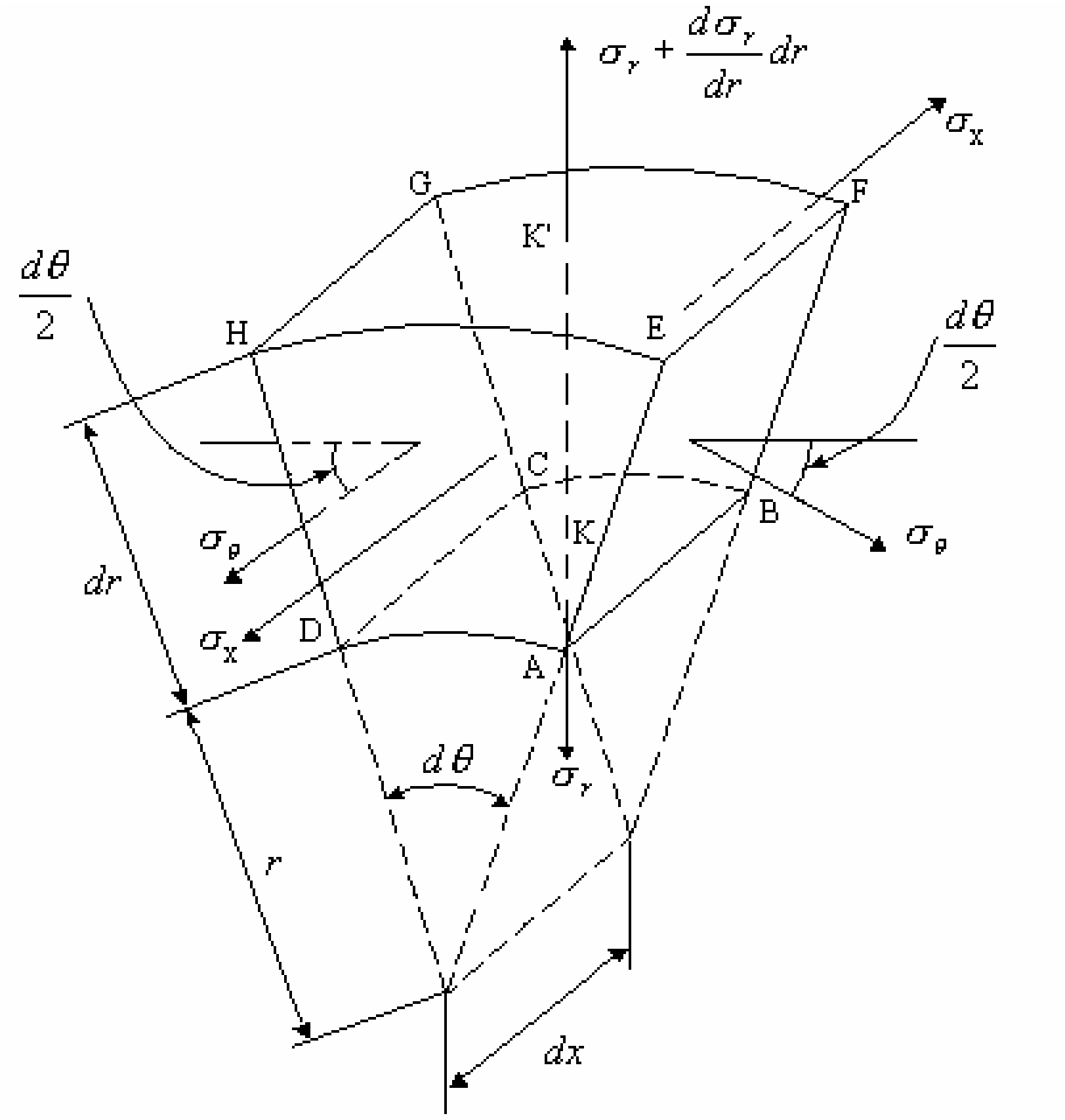

**Fig. 1** Stress distribution in the axialsymmetric cylindrical structures according to the simplified elastic theory (extracted from [1]).

By introducing cylindrical coordinates ($x$, $\theta$, $r$), the inner radius of the elementary volume is $r$ and the outer radius is $r+dr$ ($dr$ is infinitesimal). The volume is transversely limited by the surfaces *ADHE*, *BCGF*, by the surfaces *ABFE* , *CDHG* on two planes rotated of an infinitesimal angle $d\vartheta$ and by the surfaces *ABCD, EFGH* defined by the inner radius $r$ and outer radius $r+dr$ respectively.

We can consider the following nomenclature:

- $\sigma_\theta$ : circumferential stress;
- $\sigma_r$ and $\sigma_r + \frac{d\sigma_r}{dr}dr$ : radial stresses applied to the surface *ABCD* and *EFGH* respectively;
- $r \cdot d\theta \cdot dx$ and $(r+dr)\cdot d\theta \cdot dx$ : internal surface *ABCD* and the external surface *EFGH* of the elementary volume respectively;
- $dr \cdot dx$ : surface *ABEF* and *CDGH*
- $\sigma_r \cdot r \cdot d\vartheta \cdot dx$ : force applied to internal surface *ABCD* of the elementary volume;
- $\left(\sigma_r + \frac{d\sigma_r}{dr}dr\right)\cdot(r+dr)\cdot d\theta \cdot dx$ : force applied to the external surface *EFGH* of the elementary volume;
- $\sigma_\theta \cdot sen\frac{d\theta}{2} \cong \sigma_\theta \cdot \frac{d\theta}{2}$ : vertical component (parallel to the $r$ direction KK') of the circumferential stress;
- $\sigma_\theta \cdot \frac{d\theta}{2} \cdot drdx$ : vertical component of the force applied to the surfaces *ABFE* and *CDHG*.
- $E$ and $\nu$ Young's Modulus and Poisson's ratio respectively.

The formulation of the *The Lamé Problem* is based on the following hypotheses:

i) axial-symmetry: $u$ (in $r$ direction) is the only displacement component;

ii) radial displacement $u$ only depend on the radius $r$:

$$u(r,x,\theta) = u(r)$$

iii) shear stresses on the elementary volume must be zero: $\tau_{xr} = \tau_{x\theta} = \tau_{r\theta} = 0$;

iv) on the basis of the axial-symmetry and constant thickness, the radial and circumferential stresses only depend on the radius $r$:

$$\begin{cases} \sigma_r(r,x,\theta) = \sigma_r(r) \\ \sigma_\theta(r,x,\theta) = \sigma_\theta(r) \end{cases}$$

v) it is assumed that the axial stress $\sigma_x$ s constant along the length of the transversal section (independent from the coordinates $x$, $r$):

$$\frac{\partial \sigma_x}{\partial r} = \frac{\partial \sigma_x}{\partial x} = 0$$

vi) axial symmetrical loads: because nothing varies in the $\theta$ direction,

vii) the problem is solved by linear-elastic approach.

Now considering the radial force equilibrium of the element in fig.1:

$$\left(\sigma_r + \frac{d\sigma_r}{dr}dr\right)\cdot(r+dr)d\theta \cdot dx - \sigma_r \cdot rd\theta \cdot dx - 2\sigma_\theta dxdr \cdot sen\left(\frac{d\theta}{2}\right) \cong$$

$$\cong \left(\sigma_r + \frac{d\sigma_r}{dr}dr\right)\cdot(r+dr)d\theta \cdot dx - \sigma_r \cdot rd\theta \cdot dx - 2\sigma_\theta dxdr\frac{d\theta}{2} =$$

$$= \sigma_r r d\theta \cdot dx + \sigma_r dr \cdot d\theta \cdot dx + \frac{d\sigma_r}{dr} dr^2 d\theta \cdot dx + \frac{d\sigma_r}{dr} r dr d\theta \cdot dx - \sigma_r r d\theta \cdot dx - \sigma_\theta dr d\theta dx =$$

$$= \sigma_r dr \cdot d\theta \cdot dx + \frac{d\sigma_r}{dr} dr^2 d\theta \cdot dx + \frac{d\sigma_r}{dr} r dr d\theta \cdot dx - \sigma_\theta dr d\theta dx = 0$$

Neglecting the product of small quantity $(d\sigma_r/dr)dr^2 d\theta\, dx$ and collecting terms, the above equation reduces to:

$$\frac{d\sigma_r}{dr} + \frac{\sigma_r - \sigma_\theta}{r} = 0 \qquad (1)$$

In order to solve the problem there is one equation with two variables $\sigma_r$ ,$\sigma_\theta$. We can determine the second equation applying the Strain-Displacements Equations:

$$\varepsilon_r = \frac{du}{dr}$$
$$\varepsilon_\theta = \frac{2\pi \cdot (r+u) - 2\pi r}{2\pi r} = \frac{u}{r} \qquad (2)$$

From Hooke’s generalized law we can obtain the following equations:

$$\varepsilon_r = \frac{1}{E} \cdot [\sigma_r - \nu \cdot (\sigma_x + \sigma_\theta)]$$
$$\varepsilon_\theta = \frac{1}{E} \cdot [\sigma_\theta - \nu \cdot (\sigma_r + \sigma_x)] \qquad (3)$$

And from the first equation of (3) we can obtain:

$$\sigma_r = E \cdot \varepsilon_r + \nu \cdot (\sigma_x + \sigma_\theta) \qquad (4)$$

Substituting eq. (4) in the second equation of (3) we get:

$$\varepsilon_\theta = \frac{1}{E} \cdot [\sigma_\theta - \nu \cdot E\varepsilon_r - \nu^2 \cdot \sigma_x - \nu^2 \cdot \sigma_\theta - \nu \cdot \sigma_x] = \frac{1}{E} \cdot [\sigma_\theta \cdot (1-\nu^2) - \nu\sigma_x \cdot (1+\nu) - \nu E \cdot \varepsilon_r]$$

$$E\varepsilon_\theta = \sigma_\theta \cdot (1-\nu^2) - \nu\sigma_x \cdot (1+\nu) - \nu E \cdot \varepsilon_r \quad \Rightarrow \quad E\varepsilon_\theta + \nu\sigma_x \cdot (1+\nu) + \nu E \cdot \varepsilon_r = \sigma_\theta \cdot (1-\nu^2)$$

$$\sigma_\theta = \frac{1}{1-\nu^2} \cdot [E\varepsilon_\theta + \nu\sigma_x \cdot (1+\nu) + \nu E\varepsilon_r]$$

and collecting terms:

$$\sigma_\theta = \frac{E}{1-\nu^2} \cdot (\varepsilon_\theta + \nu\varepsilon_r) + \frac{\nu\sigma_x}{1-\nu} \qquad (5)$$

Substituting eq. (5) in eq. (4):

$$\sigma_r = E\varepsilon_r + \nu\sigma_x + \frac{\nu E}{1-\nu^2} \cdot (\varepsilon_\theta + \nu\varepsilon_r) + \frac{\nu^2\sigma_x}{1-\nu} = \frac{E\varepsilon_r(1-\nu^2) + \nu E \cdot (\varepsilon_\theta + \nu\varepsilon_r)}{1-\nu^2} + \frac{\nu\sigma_x(1-\nu) + \nu^2\sigma_x}{1-\nu} =$$
$$= \frac{E\varepsilon_r - E\varepsilon_r\nu^2 + \nu E\varepsilon_\theta + E\varepsilon_r\nu^2}{1-\nu^2} + \frac{\nu\sigma_x - \nu^2\sigma_x + \nu^2\sigma_x}{1-\nu}$$

With opportune simplifications and collecting terms:

$$\sigma_r = \frac{E}{1-\nu^2} \cdot (\varepsilon_r + \nu\varepsilon_\theta) + \frac{\nu\sigma_x}{1-\nu} \qquad (6)$$

Substituting the equations (2) in (5) and (6), the following stress-displacements equations can be obtained:

$$\sigma_\theta = \frac{E}{1-\nu^2}\cdot\left(\frac{u}{r}+\nu\frac{du}{dr}\right)+\frac{\nu\sigma_x}{1-\nu}$$
$$\sigma_r = \frac{E}{1-\nu^2}\cdot\left(\frac{du}{dr}+\nu\frac{u}{r}\right)+\frac{\nu\sigma_x}{1-\nu} \quad (7)$$

Substituting eqs. (7) in eq. (1) and with opportune simplifications, we can obtain:

$$\frac{E}{1-\nu^2}\cdot\frac{d}{dr}\left(\frac{du}{dr}+\nu\frac{u}{r}\right)+\frac{E}{1-\nu^2}\cdot\frac{1}{r}\left(\frac{du}{dr}+\nu\frac{u}{r}-\frac{u}{r}-\nu\frac{du}{dr}\right)=0$$

$$\left(\frac{d^2u}{dr^2}+\frac{\nu}{r}\frac{du}{dr}-\nu\frac{u}{r^2}\right)+\left(\frac{1}{r}\frac{du}{dr}+\nu\frac{u}{r^2}-\frac{u}{r^2}-\frac{\nu}{r}\frac{du}{dr}\right)=0$$

$$\frac{d^2u}{dr^2}+\frac{\nu}{r}\frac{du}{dr}-\nu\frac{u}{r^2}+\frac{1}{r}\frac{du}{dr}+\nu\frac{u}{r^2}-\frac{u}{r^2}-\frac{\nu}{r}\frac{du}{dr}=0$$

And with opportune simplifications:

$$\frac{d^2u}{dr^2}+\frac{1}{r}\frac{du}{dr}-\frac{u}{r^2}=0$$

Which can be expressed as:

$$\frac{d}{dr}\left(\frac{du}{dr}+\frac{u}{r}\right)=0 \quad (8)$$

Furthermore eq. (8) can be written as:

$$\frac{d}{dr}\left(\frac{du}{dr}+\frac{u}{r}\right)=\frac{d}{dr}\left[\frac{1}{r}\left(r\cdot\frac{du}{dr}+u\right)\right]=\frac{d}{dr}\left[\frac{1}{r}\left(r\cdot\frac{du}{dr}+\frac{dr}{dr}\cdot u\right)\right]=0$$

And finally a more compact form of eq. (8) is obtained:

$$\frac{d}{dr}\left[\frac{1}{r}\frac{d}{dr}(u\cdot r)\right]=0 \quad (9)$$

Integrating this equation:

$$\frac{1}{r}\frac{d}{dr}(u\cdot r)=C_0$$

$$u\cdot r=C_0\cdot\frac{r^2}{2}+C_2$$

$$u=C_0\cdot\frac{r}{2}+\frac{C_2}{r}$$

By assuming: $C_0/2 = C_1$, we can obtain the radial displacements as a function of the radius:

$$u=C_1\cdot r+\frac{C_2}{r} \quad \text{defined for } r\neq 0 \quad (10)$$

Substituting eq. (10 ) in stress-displacement equations (7) we can determine the circumferential and the radial stresses at any point through the thickness of the wall.

In order to determine the integration constants $C_1$ and $C_2$, opportune boundary conditions are necessary.

## 3. Controlled clearance pressure balances with cylinder made up of single material

In the case of CCPGs the solution of The *Lamé Problem* is based on the following model adopted for the schematization of the CC PCU(fig.2):

I. The thick-walled cylinder with open ends (inner radius $r_c$ and outer radius $R_c$) is subject to internal pressure $p$ and external jacket pressure $p_j$ (fig.2) where $p$ is the pressure distribution in the clearance.
II. no end-loading on cylinder: $\sigma_x = 0$ [6]
III. the piston is subject to axial stress $\sigma_x$ = cost = $-P$ and external pressure $p$. Where P is the pressure measurement applied to the piston base.

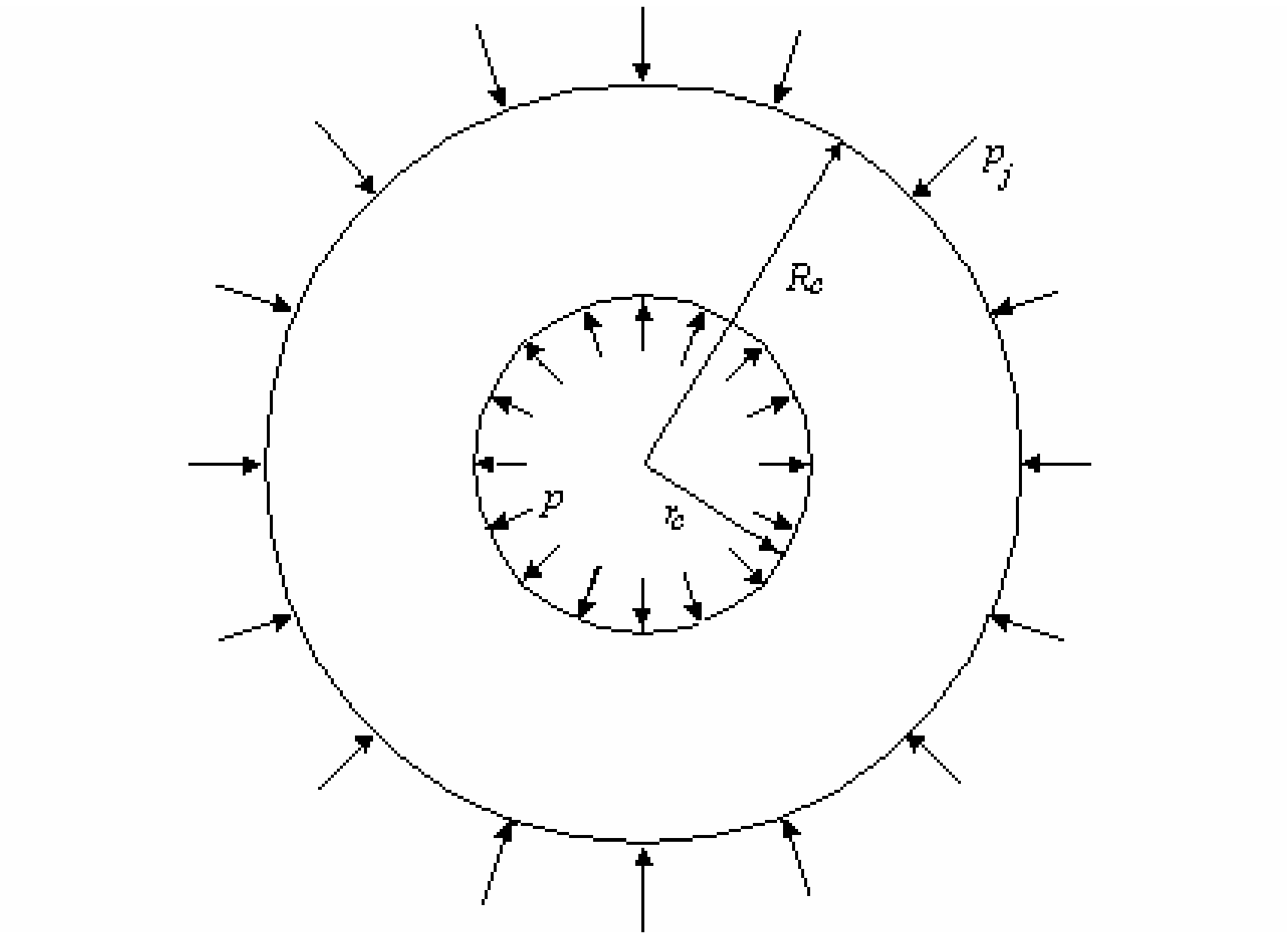

**Fig.2** Transversal section of the cylinder (constituted by single material) of a controlled-clearance pressure balance (extracted from [1]).

Eq. (10) allows us to determine the parametric equation of the strains [5]:

$$\frac{du}{dr} = C_1 - \frac{C_2}{r^2} = \varepsilon_r$$
$$\frac{u}{r} = C_1 + \frac{C_2}{r^2} = \varepsilon_\theta \qquad (11)$$

Substituting eqs. (11) in stress-displacement equations (7) we can obtain:

$$\sigma_r(r) = \frac{E_c}{1-\nu_c^{\,2}} \cdot \left[ C_1 - \frac{C_2}{r^2} + \nu_c \cdot \left( C_1 + \frac{C_2}{r^2} \right) \right] + \frac{\nu_c \sigma_x}{1-\nu_c}$$

$$\sigma_\theta(r) = \frac{E_c}{1-\nu_c^{\,2}} \cdot \left[ C_1 + \frac{C_2}{r^2} + \nu_c \cdot \left( C_1 - \frac{C_2}{r^2} \right) \right] + \frac{\nu_c \sigma_x}{1-\nu_c}$$

With opportune changes the following equations can be obtained:

$$\sigma_r(r) = \frac{E_c}{1-\nu_c^{\,2}} \cdot \left[ C_1 \cdot (1+\nu_c) - C_2 \cdot (1-\nu_c) \frac{1}{r^2} \right] + \frac{\nu_c \sigma_x}{1-\nu_c}$$

$$\sigma_\theta(r) = \frac{E_c}{1-\nu_c^{\,2}} \cdot \left[ C_1 \cdot (1+\nu_c) + C_2 \cdot (1-\nu_c) \frac{1}{r^2} \right] + \frac{\nu_c \sigma_x}{1-\nu_c} \qquad (12)$$

And for the analysis of the cylinder: $r_c \leq r \leq R_c$.

Where $E_c$ and $\nu_c$ are the Young's modulus and Poisson's ratio of the cylinder respectively.

*3.1 Boundary conditions*

The following boundary conditions allow us to determine the integration constants $C_1$ and $C_2$:

$$\begin{cases} \sigma_r(R_c) = \dfrac{E_c}{1-\nu_c^2} \cdot \left[ C_1 \cdot (1+\nu_c) - C_2 \cdot (1-\nu_c) \dfrac{1}{R_c^2} \right] + \dfrac{\nu_c \sigma_x}{1-\nu_c} = -p_j \\ \\ \sigma_r(r_c) = \dfrac{E_c}{1-\nu_c^2} \cdot \left[ C_1 \cdot (1+\nu_c) - C_2 \cdot (1-\nu_c) \dfrac{1}{r_c^2} \right] + \dfrac{\nu_c \sigma_x}{1-\nu_c} = -p \end{cases} \tag{13}$$

The solution of the system (13) is carried out with the following difference:

$$\sigma_r(R_c) - \sigma_r(r_c) = \frac{E_c}{1-\nu_c^2} \cdot \left[ -C_2 \cdot (1-\nu_c) \frac{1}{R_c^2} + C_2 \cdot (1-\nu_c) \frac{1}{r_c^2} \right] = -p_j + p$$

$$-C_2 \cdot (1-\nu_c) \frac{1}{R_c^2} + C_2 \cdot (1-\nu_c) \frac{1}{r_c^2} = (p - p_j) \cdot \left( \frac{1-\nu_c^2}{E_c} \right)$$

$$C_2 \cdot \left[ \frac{(R_c^2 - r_c^2) \cdot (1-\nu_c)}{r_c^2 \cdot R_c^2} \right] = (p - p_j) \cdot \left( \frac{1-\nu_c^2}{E_c} \right)$$

The first integration constant is equal to:

$$C_2 = (p - p_j) \cdot \left( \frac{1-\nu_c^2}{E_c} \right) \cdot \left[ \frac{r_c^2 \cdot R_c^2}{(R_c^2 - r_c^2) \cdot (1-\nu_c)} \right] \tag{14}$$

Substituting eq. (14) in the second eq. in (13) we can obtain:

$$\frac{E_c}{1-\nu_c^2} \cdot \left[ C_1 \cdot (1+\nu_c) - \frac{(p - p_j) \cdot (1-\nu_c^2) \cdot r_c^2 \cdot R_c^2}{E_c \cdot (R_c^2 - r_c^2) \cdot (1-\nu_c)} \cdot (1-\nu_c) \frac{1}{r_c^2} \right] = -p - \frac{\nu_c \sigma_x}{1-\nu_c}$$

$$\frac{E_c}{1-\nu_c^2} \cdot \left[ C_1 \cdot (1+\nu_c) - \frac{(p - p_j) \cdot (1-\nu_c^2) \cdot R_c^2}{E_c \cdot (R_c^2 - r_c^2)} \right] = -p - \frac{\nu_c \sigma_x}{1-\nu_c}$$

$$\frac{C_1 \cdot (1+\nu_c) \cdot E_c}{1-\nu_c^2} - \frac{(p - p_j) \cdot R_c^2}{(R_c^2 - r_c^2)} = -p - \frac{\nu_c \sigma_x}{1-\nu_c}$$

$$C_1 = \left[ \frac{(p - p_j) \cdot R_c^2 - p \cdot (R_c^2 - r_c^2)}{(R_c^2 - r_c^2)} \right] \cdot \frac{1-\nu_c}{E_c} - \frac{\nu_c \sigma_x}{E_c} = \left[ \frac{p \cdot R_c^2 - p_j \cdot R_c^2 - p \cdot R_c^2 + p \cdot r_c^2}{(R_c^2 - r_c^2)} \right] \cdot \frac{1-\nu_c}{E_c} - \frac{\nu_c \sigma_x}{E_c}$$

The second integration constant is equal to:

$$C_1 = \left[ \frac{p \cdot r_c^2 - p_j \cdot R_c^2}{(R_c^2 - r_c^2)} \right] \cdot \frac{1-\nu_c}{E_c} - \frac{\nu_c \sigma_x}{E_c} \tag{15}$$

Substituting eqs. (14) and (15) in eq. (10) we can obtain:

$$U(r) = \left[ \frac{p \cdot r_c^2 - p_j \cdot R_c^2}{(R_c^2 - r_c^2)} \cdot \frac{1-\nu_c}{E_c} - \frac{\nu_c \sigma_x}{E_c} \right] r + \left( \frac{p - p_j}{r} \right) \cdot \left( \frac{1-\nu_c^2}{E_c} \right) \cdot \left[ \frac{r_c^2 \cdot R_c^2}{(R_c^2 - r_c^2) \cdot (1-\nu_c)} \right]$$

We can determine the radial displacements of the cylinder (as a function of *r*) with the following equation:

$$U(r)=\frac{1-\nu_c}{E_c}\left(\frac{pr_c^2-p_jR_c^2}{R_c^2-r_c^2}\right)r+\frac{1+\nu_c}{E_c}\left(\frac{r_c^2R_c^2}{r}\right)\frac{p-p_j}{R_c^2-r_c^2}-\frac{\nu_c}{E_c}\sigma_x r \tag{16}$$

Remembering the hypothesis of no end-loading on cylinder ($\sigma_x = 0$) we can obtain:

$$U(r)=\frac{1-\nu_c}{E_c}\left(\frac{pr_c^2-p_jR_c^2}{R_c^2-r_c^2}\right)r+\frac{1+\nu_c}{E_c}\left(\frac{r_c^2R_c^2}{r}\right)\frac{p-p_j}{R_c^2-r_c^2} \tag{17}$$

The eq. (17) is used to determine the radial cylinder displacements for $r = r_c$:

$$U(r_c)=\frac{1-\nu_c}{E_c}\left(\frac{pr_c^2-p_jR_c^2}{R_c^2-r_c^2}\right)r_c+\frac{1+\nu_c}{E_c}\left(\frac{r_c^2R_c^2}{r_c}\right)\frac{p-p_j}{R_c^2-r_c^2}=$$

$$=\frac{r_c}{E_c}\left[\frac{p\left(R_c^2+r_c^2\right)-2p_jR_c^2+\nu_c p\left(R_c^2-r_c^2\right)}{R_c^2-r_c^2}\right]$$

and finally:

$$U(r_c)=\frac{r_c p}{E_c}\left[\frac{R_c^2+r_c^2-2\left(\frac{p_j}{p}\right)R_c^2}{R_c^2-r_c^2}+\nu_c\right] \tag{18}$$

With the same procedure and remember the hypothesis III (page 6), the radial piston distortions can be obtained:

$$u(r)=-\left(\frac{1-\nu_p}{E_p}\right)p\cdot r-\frac{\nu_c}{E_c}\sigma_x r$$

By assuming $\sigma_x = -P$ we can obtain the radial piston distortions as a function of radius and pressure distribution in the clearance *p*:

$$u(r)=-\left(\frac{1-\nu_p}{E_p}\right)p\cdot r+\left(\frac{\nu_p\cdot P}{E_p}\right)r \tag{19}$$

The eq. (19) is used to determine the radial piston distortions for $r = r_p$:

$$u(r_p)=\frac{r_p}{E_p}\left[\left(\nu_p-1\right)p+\nu_p P\right] \tag{20}$$

Where $E_p$ and $\nu_p$ are the Young's Modulus and Poisson's ratio of the piston respectively.

### *3.2 Pressure distortion coefficient*

The pressure distortion coefficient of a controlled-clearance pressure balance is carried out with the equations (18) and (20). The evaluation of the distortion coefficient is made with the same hypotheses formulated in [6] (for simple and reentrant units):

i) constant gap between the piston and the cylinder;
ii) the dimensions of the clearance are negligible; the radius of the piston is equal to the internal radius of the cylinder: $r_c \cong r_p$ and $h_0 \cong 0$; where $h_0$ is the undistorted gap between the piston and the cylinder;
iii) linear pressure profile in the clearance along the engagement piston-cylinder length *L*: in this analysis we can assume an average constant pressure profile with pressure value equal to *0,5 P*.

According to the hypothesis ii) page 3):

$$\frac{dU}{dx} = \frac{du}{dx} = 0 \;\Rightarrow \begin{cases} u_0 = u_L \\ U_0 = U_L \end{cases}$$

where $u_0$ and $U_0$ are the piston and cylinder displacements at the gap entrance respectively while $u_L$ and $U_L$ are the piston and cylinder displacements at the gap exit respectively.
Remembering the equation of the pressure distortion coefficient (Pavese and Molinar 1992):

$$\lambda = \frac{1}{1+\frac{h_0}{r_p}} \cdot \left[ \frac{u_0 + U_0}{P \cdot r_p} + \frac{1}{P^2 \cdot r_p} \int_0^l p(x) \cdot \left( \frac{dU}{dx} + \frac{du}{dx} \right) dx \right]$$

and for the hypotheses adopted we can obtain:

$$\lambda = \frac{1}{1+\frac{h_0}{r_p}} \cdot \left( \frac{u_0 + U_0}{P \cdot r_p} \right) \cong \frac{u_0 + U_0}{P \cdot r_c} \tag{21}$$

Remembering the hypotheses made for this model (pag. 3) and substituting $p = P/2$ in eqs. (18) and (20), the radial displacements of the cylinder and the radial displacements of piston are as follows.

$$U_0 = U_L = \frac{r_c P}{2E_c} \left[ \frac{\left(R_c^{\,2} + r_c^2\right) - \frac{4p_j}{P}}{R_c^{\,2} - r_c^{\,2}} + \nu_c \right] = \frac{r_c P}{2E_c} \left[ \frac{r_c^2 + R_c^{\,2} - 4t \cdot R_c^2}{R_c^{\,2} - r_c^{\,2}} + \nu_c \right]$$

(22)

$$u_0 = u_L = \frac{r_p}{E_p} \left[ \nu_p \frac{P}{2} - \frac{P}{2} + \nu_p P \right] \cong \frac{r_c P}{2E_p} \left(3\nu_p - 1\right)$$

With the ratio: $t = p_j/P$. Substituting eqs. (22) in the eq. (21) we can obtain:

$$\lambda = \frac{1}{P \cdot r_c} \cdot \frac{r_c P}{2E_p} \left(3\nu_p - 1\right) + \frac{1}{P \cdot r_c} \cdot \frac{r_c P}{2E_c} \left( \frac{r_c^2 + R_c^{\,2} - 4t \cdot R_c^2}{R_c^{\,2} - r_c^{\,2}} + \nu_c \right)$$

With opportune simplifications we can determine the pressure distortion coefficient for controlled-clearance pressure balances[1]:

$$\lambda = \frac{\left(3\nu_p - 1\right)}{2E_p} + \frac{1}{2E_c} \left( \frac{R_c^{\,2} + r_c^2 - 4t \cdot R_c^2}{R_c^{\,2} - r_c^{\,2}} + \nu_c \right) \tag{23}$$

By assuming $p_j = 0$ in eq. (23), the pressure distortion coefficient for pressure balance in the free-deformation mode can be obtained:

$$\lambda = \frac{\left(3\nu_p - 1\right)}{2E_p} + \frac{1}{2E_c} \left( \frac{R_c^{\,2} + r_c^2}{R_c^{\,2} - r_c^{\,2}} + \nu_c \right) \tag{24}$$

Equation (24) is the formula given in [6] for the case o free deformation mode.
Eq. (23) represents a general formula (CCPGs). Thus eq. (24) represents a particular case (pressure balances in free deformation mode) of CCPGs.

The ratio $t$ was successively substituted in eq. (23) (by other authors [2][3]) with the ratio $\chi = p_j/\bar{p}$ and assuming $\bar{p} = P/2$: in that case the following formula can be obtained (Buonanno G. *et al.*, 2003):

$$\lambda = \frac{(3\nu_p - 1)}{2E_p} + \frac{1}{2E_c}\left(\frac{R_c^{\,2} + r_c^2 - 2\chi \cdot R_c^2}{R_c^{\,2} - r_c^{\,2}} + \nu_c\right) \qquad (25)$$

but this is the same formula reported in[1]: equation (23) with different notations; in fact the application of this formula gives the same numerical values.

## 4. Controlled clearance pressure balances with the cylinder made up of two materials

The technology based on the axial symmetric structures with the cylinder made up of two materials is used in the construction of pressure balances for high pressure measurements. The analysis of this configuration is based on the same hypotheses formulated in the case of the cylinder made up of single material:

I) the thick-walled cylinder 1 with open ends, is subject to internal pressure $p$ and external pressure $p_m$ (fig.4a) where $p$ is the pressure distribution in the clearance between the piston and the cylinder. The thick-walled cylinder 2 with open ends is subject to internal pressure $p_m$ and external jacket pressure $p_j$ (fig.4b).

II) no end-loading on cylinder: $\sigma_x = 0$

III) piston loaded with a longitudinal stress $\sigma_x$ = cost = $-P$ and an external pressure $p$.

We can assume the following notation(figures 3, 4):

- $R_c$: radius of the external surface of the cylinder 2;
- $r_c$: radius of the internal surface of the cylinder 1;
- $p$: pressure applied to the internal surface of cylinder 1: pressure distribution in the clearance between piston and cylinder.
- $p_j$: pressure applied to the external surface of cylinder 2: jacket pressure.
- $r_m$: radius of the external surface of the cylinder 1 equal to the radius of the internal surface of the cylinder 2;
- $E_1, \nu_1$: Young's modulus and Poisson's ratio of the internal cylinder;
- $E_2, \nu_2$: Young's modulus and Poisson's ratio of the external cylinder;
- $U_1(r)$: radial elastic displacements of the internal cylinder 1.
- $U_2(r)$:: radial elastic displacements of the external cylinder 2.
- $p_m$: interface pressure;
- $P$: pressure applied to the piston base: pressure measurement.

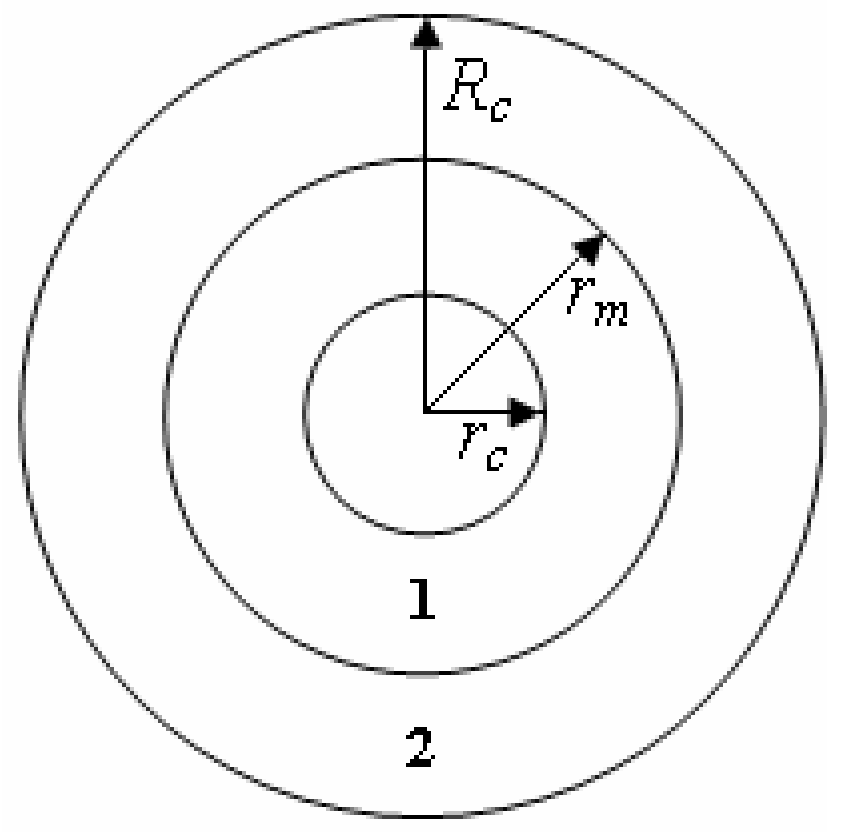

**Fig. 3** Transversal section of the cylinder made up of two materials for a controlled-clearance pressure Balance.

Also for the CCPGs with the cylinder made up of two materials, the evaluation of the pressure distortion coefficient $\lambda$ is based on the solution of the *Lamé Problem*. For this theoretical analysis the literature gives two models:

- MODEL I (Saragosa S., 2001).

A model without the external cylinder (fig.4a): the effect of cylinder 2 is replaced by the pressure $p_m$ applied to the external surface of the internal cylinder (cylinder 1); the assembly can be schematized as a controlled-clearance pressure balance with the cylinder made up of single material: the jacket pressure is $p_m$ (interface pressure). In this particular case the formula of the pressure distortion coefficient derives from the equation (23) [1]:

$$\lambda = \frac{(3\nu_p - 1)}{2E_p} + \frac{1}{2E_1}\left(\frac{r_c^{\,2} + r_m^2 - 4\varphi \cdot r_m^2}{r_m^{\,2} - r_c^{\,2}} + \nu_1\right) \tag{26}$$

with $\varphi = p_m/P$

- MODEL II (Buonanno G. *et al.*, 2003-2007).

A model without the internal cylinder 1 fig. 4b): it's obtained by "removing the internal layer 1 and replacing it with the contact pressure, $p_{med}$, applied on the layer 2" as reported in [2] and in pag.99 of [3]. This model is wrong for the solution of the *The Lamé Problem* because:

a) it is not a model of a pressure balance;
b) model II was erroneously correlated (by the authors of the previous publications [2,3]) to the equation (28): the wrong model for the right formula; in fact the analysis based on model II leads to another equation (completely different from eq. (28)) of the pressure distortion coefficient for the pressure balance with the cylinder made up of two materials: obviously for this model the introduction of piston is not allowed; thus the ratio $(3\nu_p - 1)/2E_p$ is not present in eqs. (26, 28).

This model is valid only for the calculation of $p_m$ (interface pressure).

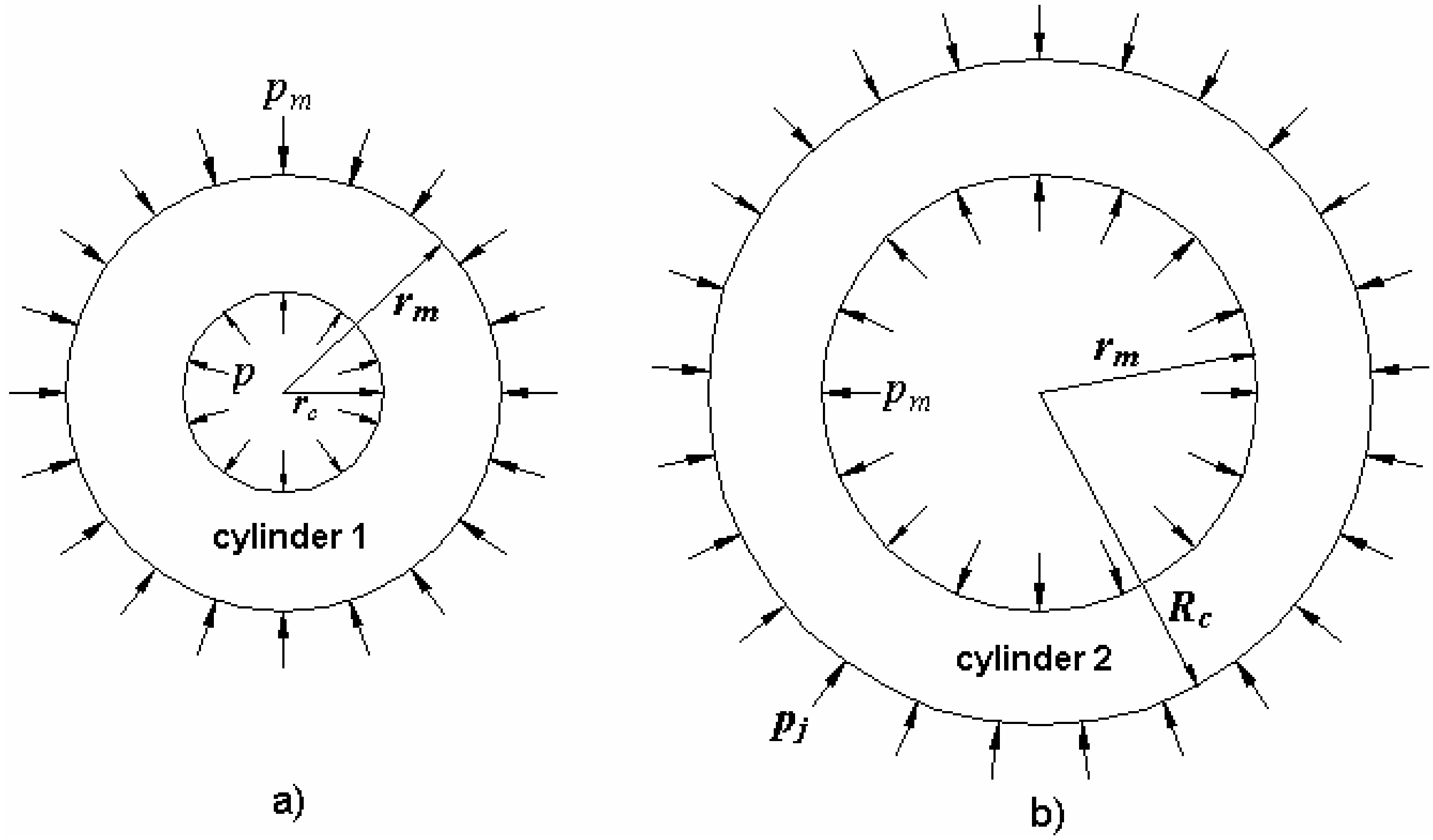

**Fig. 4** Transversal sections of the cylinder of CCPG (cylinder made of up of 2 materials) for two hypothetical models.

On the basis of the above considerations, this analysis is made using the model I.

### *4.1 Boundary conditions*

In the particular case of a CCPG with the cylinder made up of two material, the analysis is based on the evaluation of the interface pressure $p_m$ for $r = r_m$. The boundary condition for the internal cylinder 1 is:

$$U_1(r_m) = \frac{1-\nu_1}{E_1}\left(\frac{pr_c^2 - p_m r_m^2}{r_m^2 - r_c^2}\right) r_m + \frac{1+\nu_1}{E_1}\left(\frac{r_c^2 r_m^2}{r_m}\right)\frac{p - p_m}{r_m^2 - r_c^2} =$$

$$= \frac{r_m}{E_1}\left(\frac{pr_c^2 - p_m r_m^2 - p\nu_1 r_c^2 + p_m \nu_1 r_m^2 + pr_c^2 - p_m r_c^2 + pr_c^2\nu_1 - p_m r_c^2 \nu_1}{E_1(r_m^{\,2} - r_c^2)}\right) =$$

$$= \frac{r_m}{E_1}\left[\frac{p_m r_m^2(\nu_1 - 1) - p_m r_c^2(1+\nu_1) + 2pr_c^2}{(r_m^{\,2} - r_c^2)}\right]$$

and for the external cylinder 2 is:

$$U_2(r_m) = \frac{r_m}{E_2}\left[\frac{(1-\nu_2)\cdot\left(p_m r_m^2 - p_j R_c^{\,2}\right) + R_c^2(1+\nu_2)\cdot\left(p_m - p_j\right)}{R_c^2 - r_m^2}\right] =$$

$$= \frac{r_m}{E_2}\left[\frac{p_m r_m^2(1-\nu_2) + p_m R_c^2(1+\nu_2) - 2p_j R_c^2}{(R_c^2 - r_m^2)}\right]$$

For $r = r_m$ the radial elastic displacements of the points of the internal cylinder 1 are equal to radial elastic displacements of the points of the external cylinder 2:

$$U_1(r_m) = \frac{r_m}{E_1}\left[\frac{p_m r_m^2(\nu_1 - 1) - p_m r_c^2(1+\nu_1) + 2pr_c^2}{(r_m^{\,2} - r_c^2)}\right] = U_2(r_m) = \frac{r_m}{E_2}\left[\frac{p_m r_m^2(1-\nu_2) + p_m R_c^2(1+\nu_2) - 2p_j R_c^2}{(R_c^2 - r_m^2)}\right]$$

We can introduce the ratios $\beta_1^2 = \dfrac{r_c^2}{r_m^2}$, $\beta_2^2 = \dfrac{r_m^2}{R_c^2}$:

$$\frac{p_m\left[(\nu_1 - 1) - (1+\nu_1)\cdot\beta_1^2\right] + 2p\beta_1^2}{E_1(1-\beta_1^2)} = \frac{p_m\left[\beta_2^2(1-\nu_2) + (1+\nu_2)\right] - 2p_j}{E_2(1-\beta_2^2)}$$

with $n = p_j/p$:

$$\frac{2p\beta_1^2}{E_1(1-\beta_1^2)} + \frac{2p_j}{E_2(1-\beta_2^2)} = 2p\left[\frac{\beta_1^2}{E_1(1-\beta_1^2)} + \frac{n}{E_2(1-\beta_2^2)}\right] =$$
$$= \frac{p_m\left[\beta_2^2(1-\nu_2) + (1+\nu_2)\right]}{E_2(1-\beta_2^2)} - \frac{p_m\left[(\nu_1 - 1) - (1+\nu_1)\cdot\beta_1^2\right]}{E_1(1-\beta_1^2)} \qquad (27)$$

By assuming:

$$A = \frac{[(\nu_1 - 1) - (1+\nu_1)\cdot\beta_1^2]}{E_1(1-\beta_1^2)}; \quad B = \frac{[\beta_2^2(1-\nu_2) + (1+\nu_2)]}{E_2(1-\beta_2^2)}; \quad C = \frac{[\beta_1^2 \cdot E_2(1-\beta_2^2) + nE_1(1-\beta_1^2)]}{E_1 E_2(1-\beta_1^2)\cdot(1-\beta_2^2)}$$

the interface pressure can be obtained:

$$p_m = \frac{2pC}{B-A}$$

*4.2 Pressure distortion coefficient*

Substituting the above equation of $p_m$ in eq. (26) and remembering the hypotheses in section 3.2 ($p = P/2$), we can obtain:

$$\lambda = \frac{(3\nu_p - 1)}{2E_p} + \frac{1}{2E_1}\left( \frac{r_m{}^2 + r_c^2 - 4\frac{p_m r_m^2}{P}}{r_m{}^2 - r_c{}^2} + \nu_1 \right) = \frac{(3\nu_p - 1)}{2E_p} + \frac{1}{2E_1}\left( \frac{r_m{}^2 + r_c^2 - 4\frac{2\frac{P}{2}Cr_m^2}{P(B-A)}}{r_m{}^2 - r_c{}^2} + \nu_1 \right)$$

And finally we can determine the pressure distortion coefficient for a controlled-clearance pressure balance in the case of cylinder made up of two materials[1]:

$$\lambda = \frac{(3\nu_p - 1)}{2E_p} + \frac{1}{2E_1}\left( \frac{r_m{}^2 + r_c^2 - 4\frac{Cr_m^2}{(B-A)}}{r_m{}^2 - r_c{}^2} + \nu_1 \right) \tag{28}$$

this equation was successively reported in [2] and [3] and used in [8,9]. From eq. (27) we can obtain:

$$2p\left[\frac{\beta_1^2}{E_1(1-\beta_1^2)} + \frac{n}{E_2(1-\beta_2^2)}\right] = \frac{p_m\left[\beta_2^2(1-\nu_2) + (1+\nu_2)\right]}{E_2(1-\beta_2^2)} + \frac{p_m\left[(1-\nu_1) + (1+\nu_1)\cdot\beta_1^2\right]}{E_1(1-\beta_1^2)} =$$

$$= p_m\left[\frac{\beta_2^2(1-\nu_2) + (1+\nu_2)}{E_2(1-\beta_2^2)} + \frac{(1-\nu_1) + (1+\nu_1)\cdot\beta_1^2}{E_1(1-\beta_1^2)}\right]$$

By assuming:

$$a = \frac{\beta_1^2}{E_1(1-\beta_1^2)} + \frac{n}{E_2(1-\beta_2^2)}$$

$$b = \frac{\beta_2^2(1-\nu_2) + (1+\nu_2)}{E_2(1-\beta_2^2)} + \frac{\beta_1^2(1+\nu_1) + (1-\nu_1)}{E_1(1-\beta_1^2)}$$

the interface pressure is equal to:

$$p_m = \frac{2pa}{b}$$

For $p = P/2$:

$$p_m = \frac{2pa}{b} = \frac{2\cdot\frac{P}{2}\cdot a}{b} = \frac{P\cdot a}{b}$$

and finally with $\varphi = p_m/P = (P\cdot a)/(b\cdot P) = a/b$ from eq. (26) we can obtain:

$$\lambda_{CC} = \frac{(3\nu_p - 1)}{2E_p} + \frac{1}{2E_1}\left( \frac{r_m{}^2 + r_c^2 - \frac{4ar_m^2}{b}}{r_m{}^2 - r_c{}^2} + \nu_1 \right) \tag{29}$$

Eq. (29) represents the same formula (more compact) given by the eq. (28). In the particular case of free deformation mode by assuming $p_j = 0$ in eq. (29) the following constants can be obtained:

$$f = \frac{\beta_1^2}{E_1(1-\beta_1^2)}$$

$$b = \frac{\beta_2^2(1-\nu_2)+(1+\nu_2)}{E_2(1-\beta_2^2)} + \frac{\cdot\,\beta_1^2(1+\nu_1)+(1-\nu_1)}{E_1(1-\beta_1^2)}$$

Thus the pressure distortion coefficient for a pressure balance in free deformation mode with cylinder made up of two materials can be also evaluated with the following formula:

$$\lambda_{FD} = \frac{(3\nu_p - 1)}{2E_p} + \frac{1}{2E_1}\left(\frac{r_m^{\ 2} + r_c^2 - \dfrac{4f\cdot r_m^2}{b}}{r_m^{\ 2} - r_c^{\ 2}} + \nu_1\right) \tag{30}$$

with subscripts "*CC*" and "*FD*" denoting controlled clearance and free deformation, respectively. From eqs. (29) and (30) is possible to derive the equation of jacket pressure distortion coefficient $n_j$ for the CC PCUs wherecylinder consists of two materials: in this regard, the analysis begins from the standard method given in[7, 8] based on the previous evaluation of $\lambda_{FD}$ and $\lambda_{CC}$:

$$\lambda_{FD} - \lambda_{CC} = n_j \cdot t \tag{31}$$

Using eqs. (29), (30) we get:

$$\lambda_{FD} - \lambda_{CC} = \frac{1}{2E_1}\left[\frac{\dfrac{4nr_m^2}{E_2(1-\beta_2^2)}}{b(r_m^{\ 2} - r_c^{\ 2})}\right] = \frac{2n}{E_1E_2b(1-\beta_1^2)(1-\beta_2^2)} =$$

$$= \frac{2n}{E_1(1-\beta_1^2)[\beta_2^2(1-\nu_2)+(1+\nu_2)] + E_2(1-\beta_2^2)[\beta_1^2(1+\nu_1)+(1-\nu_1)]}$$

and finally for $p = P/2$:

$$\lambda_{FD} - \lambda_{CC} = \frac{4(p_j/P)}{E_1(1-\beta_1^2)[\beta_2^2(1-\nu_2)+(1+\nu_2)] + E_2(1-\beta_2^2)[\beta_1^2(1+\nu_1)+(1-\nu_1)]}$$

For $t = p_j/P$ and with opportune simplifications, the following formula can be obtained:

$$n_j = \frac{(\lambda_{FD} - \lambda_{CC})}{t} = \frac{4}{E_1(1-\beta_1^2)[\beta_2^2(1-\nu_2)+(1+\nu_2)] + E_2(1-\beta_2^2)[\beta_1^2(1+\nu_1)+(1-\nu_1)]} \tag{32}$$

## 5. Comparison of methods

The results of the simplified analytical method are compared with experimental measurements and with results of Finite Elements analysis Methods (FEM) made at Physikalisch-Technische Bundesanstalt, Braunschweig, Germany (PTB), Bureau National de Métrologie, Laboratoire National d'Essais, Paris, France (BNM-LNE) and Ulusal Metroloji Enstitüsü, Gebze/Kocaeli, Turkey (UME) for pressure balances BNM-LNE 200 Mpa and PTB 1GPa DH 7594.

Up today the FEM analysis represents a widely adopted theoretical method for the characterization of pressure balances: the results of FEM simulations can be used to design assemblies with good performances.

*5.1 Pressure Balance BNM-LNE 200 MPa*

The controlled-clearance pressure balance BNM-LNE 200 MPa (operating up to 200MPa) characterized in the Euromet (now EURAMET) Project 256 with the assemblies unit #4 and unit #5(fig.2) [7] is described in table1.

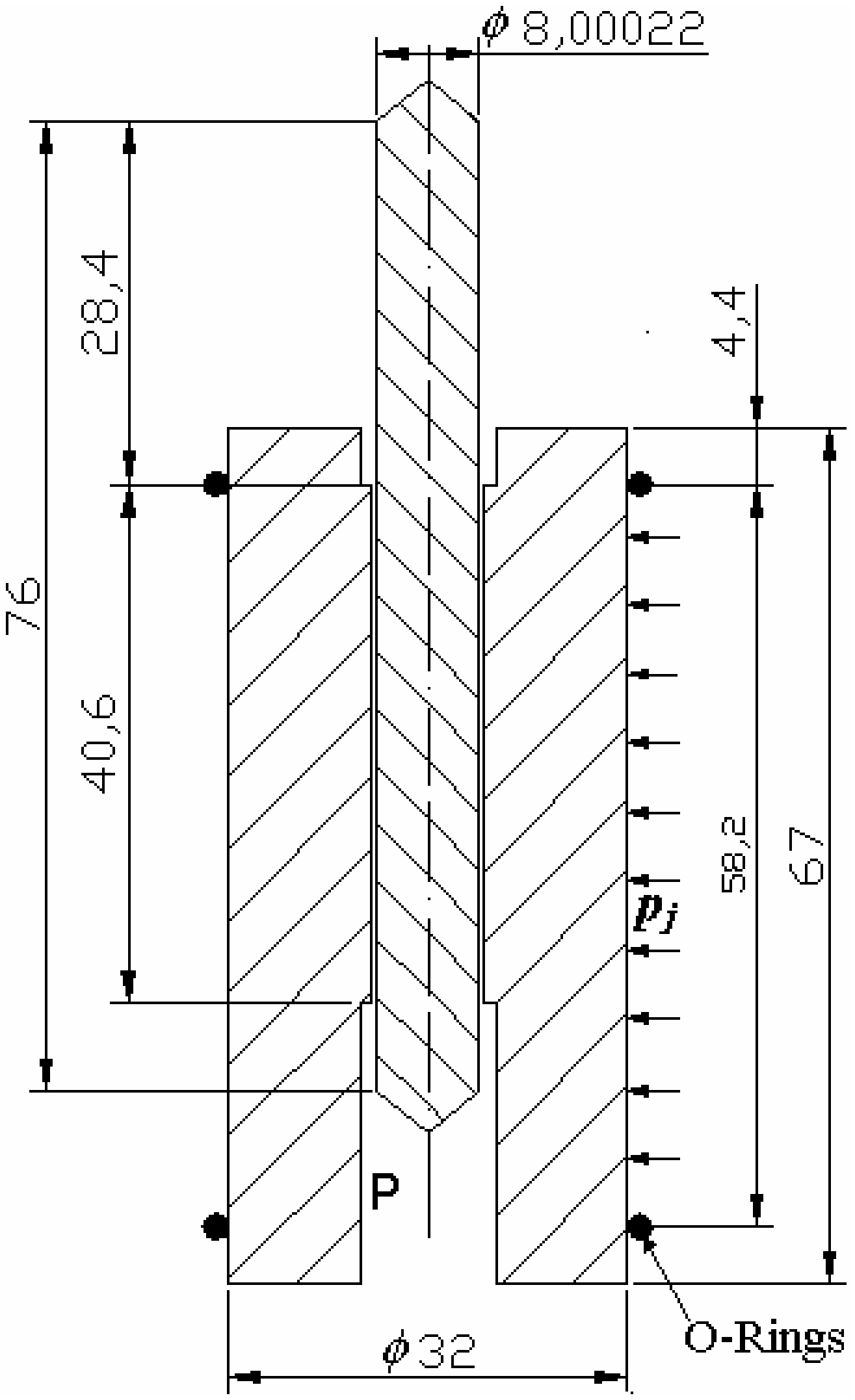

**Fig.5** BNM-LNE 200 Mpa assembly unit #5 redrawn by the author (not to scale)

**Table 1.** Main characteristics of BNM-LNE 200 Mpa assembly unit #5; those for unit #4 are given in parentheses (extracted from [7] ).

| Property | Value |
|---|---|
| Geometry | Simple/Controlled clearance |
| Pressure scale (MPa) | From 6 to 200 |
| Piston radius (mm) | 4,00011 (3,99995) |
| Cylinder radius $r_c$ (mm) | 4,00036 (4,00052) |
| Cylinder radius $R_c$ (mm) | 16 |
| Undistorted clearance(µm) | 0,25 (0,57) |
| Material (piston and cylinder) | Tungsten carbide |
| Young's modulus (MPa) | 630000 |
| Poisson's ratio | 0,218 |
| Clearance length (mm) | 40,6 |
| $p_j/P$ (controlled-clearance mode) | 1/4 |

In the case of the controlled clearance mode with $t$ = 1/4 , the analytical value of pressure distortion coefficient can be obtained using eq. (23):

$$\lambda_{cc} = -0.0487 \ p.p.m. \ MPa^{-1}$$

For the FEM analysis of the PCUs in CC mode, the complexity of the model is also due the load conditions in the external surface of the cylinder between the O-rings. This region represents a critical boundary condition because the position (and deformations) of the O-rings could influence significantly the values of the pressure distortion coefficient in CC mode. Thus the FEM simulations were made starting with a basic model with the jacket pressure applied in the part of the outer surfaces of the cylinder lying between the O-rings up to the axis of the undistorted O-rings (58,2 mm). Moreover, a model (developed at the PTB[7]) with the jacket pressure acting on the whole outer surface of the cylinder was also considered.

In table 2 is reported [7] a comparison of methods for the calculation of pressure distortion coefficients of the assembly BNM-LNE 200MPa (units #4 and #5).

**Table 2** Pressure distortion coefficients of BNM-LNE 200MPa with FEM and experimental values extracted from [7].

| $P$(MPa) | method | $\lambda_{CC}$ ($p.p.m.$ Mpa$^{-1}$) | $\lambda_{FD}$ ($p.p.m.$ Mpa$^{-1}$) |
|---|---|---|---|
| **BNM-LNE 200 MPa unit #4** | | | |
| | Experimental | -0.14 | 0.780 |
| - | Lamé | -0.0487 | 0.798(**) |
| 120 | FEM-PTB | 0.093 | 0.802 |
| | FEM-NPL | 0.083 | 0.801 |
| 200 | FEM-PTB | 0.117 | 0.803 |
| | FEM-NPL | 0.111 | 0.800 |
| | FEM-PTB(*) | -0.050(*) | |
| **BNM-LNE 200 MPa unit #5** | | | |
| | Experimental | -0.02 | 0.850 |
| - | Lamé | -0.0487 | 0.798(**) |
| 5 | FEM-PTB | 0.011 | |
| 40 | FEM-PTB | 0.097 | |
| 120 | FEM-PTB | 0.154 | 0.802 |
| | FEM-NPL | 0.153 | 0.800 |
| 200 | FEM-PTB | 0.166 | 0.803 |
| | FEM-NPL | 0.165 | 0.801 |
| | FEM-PTB(*) | -0.043(*) | |

(*) model with the jacket pressure $p_j$ acting on the whole external surface of the cylinder
(**) value obtained using eq. (24)

Table 2 shows that the $\lambda_{CC}$ numerical results (obtained with FEM analysis) are larger than the experimental ones (<0) and than the CC distortion coefficient determined by the Simplified Elastic Theory (<0). In the particular case of the FEM analysis, performed with appropriate boundary conditions (external surface of the cylinder subject to the $p_j$), the results are in accordance with the Simplified Elastic Theory.

Due to a more simplified model adopted in the simulations in the FD mode, the FEM $\lambda_{FD}$ coefficients are in good agreement with those obtained according to the Simplified Elastic theory and the values determined experimentally.

For BNM-LNE 200 MPa pressure balance (especially in the CC mode) simplified analytical method gave an indication to improve the FEM simulations.

### *5.2 Pressure Balance PTB 1GPa DH-7594*

The CCPG PTB 1GPa DH-7594 (operating up to 1GPa) used in the Physikalisch-Technische Bundesanstalt (fig.6) was characterized with the numerical FEM[8]. The characteristics of pressure balance PTB 1GPa DH-7594 are reported in table 3.

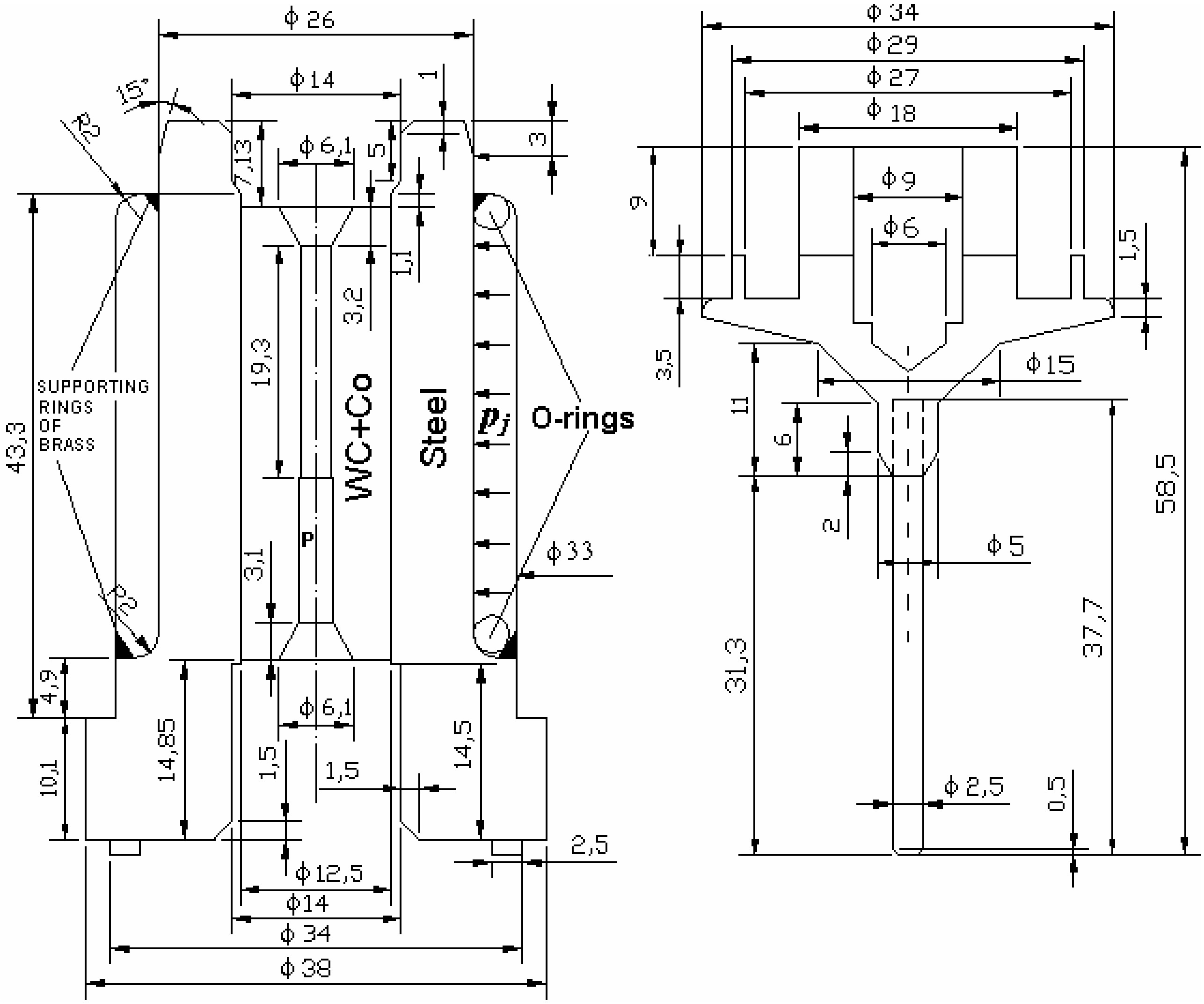

**Fig.6** Pressure balance PTB 1GPa DH-7594 redrawn by the author (not to scale).

**Table 3.** Main characteristics of PTB 1GPa DH-7594 (extracted from [8]).

| **Property** | **Value** |
|---|---|
| Geometry | Simple/Controlled clearance |
| Pressure scale (MPa) | Up to1000 (Controlled clearance mode) |
| Piston radius (mm) | 1,24899 |
| Cylinder radius $r_c$ (mm) | 1,24931 |
| Cylinder radius $r_m$ (mm) | 6,25 |
| Cylinder radius $R_c$ (mm) | 13 |
| Undistorted clearance(μm) | 0,32 |
| Material piston and inner cylinder | Tungsten carbide+Co |
| Young's modulus piston (MPa) | 543000 |
| Young's modulus inner cylinder (MPa) | 630000 |
| Young's modulus steel cylinder (MPa) | 200000 |
| Poisson's ratio (steel) | 0,29 |
| Poisson's ratio piston (carbide) | 0,238 |
| Poisson's ratio cylinder (carbide) | 0,22 |
| Axial clearance length (mm) | 19,3 |
| $p_j/P$ (controlled-clearance mode) | 1/10 |

In the FEM analysis four models were chosen [8]. One of these models (model 2) is analysed in this paper: the model with an ideal undistorted gap profile between the piston and the cylinder. The $\lambda$ calculations (with FEM) were made taking $P$ = (100, 250, 400, 600, 800, 1000) MPa in the *CC* mode and $P$ = (100, 400, 600) MPa in the *FD* mode.

By assuming $p_j/P$ = 1/10[8] (controlled clearance mode) in eq. (28) or (29), the CC pressure distortion coefficient for this assembly is equal to:

$$\lambda_{cc} = 0.354\, p.p.m.\ MPa^{-1}$$

The jacket pressure distortion coefficient can be calculated with eq.(32):

$$n_j = 3.973\, p.p.m.\ MPa^{-1}$$

From eq. (30), the pressure distortion coefficient for the case of free deformation mode can be obtained:

$$\lambda_{FD} = 0.751\, p.p.m.\ MPa^{-1}$$

The results of Simplified Elastic Theory (Lamé) are compared with FEM distortion coefficients (table 4) .

**Table 4.** some results of FEM analysis (ideal undistorted gap) and experimental values (extracted from [8]).

| $P$(MPa) | method | $\lambda_{CC}$ ($p.p.m.$ Mpa$^{-1}$) | $\lambda_{FD}$ ($p.p.m.$ Mpa$^{-1}$) | $n_j$ ($p.p.m.$ Mpa$^{-1}$) |
|---|---|---|---|---|
| **PTB 1GPa DH-7594** | | | | |
| 100 | FEM-PTB | 0.371 | 0.761 | 3.90$^{(*)}$ |
| 250 | FEM-PTB | 0.365 | 0.761 | 3.96$^{(*)}$ |
| 400 | Experimental | 0.437±0.012 | 0.776±0.011 | 3.39±0.17$^{(*)}$ |
| | FEM-PTB | 0.360 | 0.761 | 4.01$^{(*)}$ |
| | BNM-LNE | 0.335 | 0.729 | 3.94$^{(*)}$ |
| | FEM-UME | 0.370 | 0.776 | 4.06$^{(*)}$ |
| 600 | FEM-PTB | 0.356 | 0.762 | 4.06$^{(*)}$ |
| 800 | FEM-PTB | 0.354 | | |
| 1000 | FEM-PTB | 0.353 | | |
| | BNM-LNE | 0.332 | | |
| | FEM-UME | 0.356 | | |

$^{(*)}$ values obtained using equation (31) with good agreement with those obtained according to the simplified elastic theory

Table 4 shows that the application of the Simplified Elastic Theory gave results in good agreement with FEM distortion coefficients (model 2) for both the *FD* and the *CC* modes.

## 6. Conclusions

From the comparisons of methods (for two different controlled-clearance pressure balances) we can obtain the following conclusions:

- the Simplified Elastic Theory allows one to directly evaluate the pressure distortion coefficient without simulations, iterative procedures and mesh optimization (used in the FEM analysis).
- in the particular case of CCPGs where the cylinder consists of two materials, a comparison of the pressure distortion values obtained using FEM simulations (when pressure balances are modelled with an undistorted gap) together with the values obtained in the Simplified Elastic Theory reveals a good agreement as shown also in the characterizations of other controlled-clearance pressure balances in the Euromet Project 463 (pressure balance BNM-LNE 1GPa [9]);

- in order to optimize FEM simulations, sometimes the Simplified Elastic Theory represents a method for validating FEM calculations especially in the case of the deviation of FEM calculations from the experimental pressure distortion coefficients [7];

The limits of the Simplified Elastic Theory in comparison with the FEM analysis are:

- when the Simplified Theory is used, the pressure distortion coefficient uncertainties are strongly influenced by the uncertainties of the elastic constants of the materials used for the realization of the pressure balances;
- the FEM analysis is based on the numerical integration of more complex derivative equations (given in the mechanical theory of elastic equilibrium) including also the evaluation of the shear stresses: thus a more accurate analysis of the deformations of the PCUs is carried out;
- when the simulation with the FEM is undertaken, it is possible to know more detailed information about the elastic distortions in the gap region of the PCUs;
- for better understanding the real behaviour of the pressure balances (especially the more complex CCPGs) the FEM is an useful theoretical tool because more input parameters are considered in the simulation of the PCUs: for example, detailed boundary conditions and more information about the dimensional properties of the assemblies.

**REFERENCES**

[1] Saragosa Silvano, *Caratterizzazione di bilance di pressione mediante l'impiego di metodi numerici e analitici*, Master Degree Thesis, Department of Mechanical Engineering, University of Cassino-Italy, July 2001.

[2] Buonanno G, Dell'Isola M, Frattolillo A, *Simplified analytical methods to evaluate the pressure distortion coefficient of controlled-clearance pressure balances* (High Temperatures-High Pressures), 2003.

[3] Buonanno G. et al., *Ten years of experiences in modelling pressure balances in liquid media up to few GPa* (University of Cassino), February 2007.

[4] Dadson R. S., Lewis S. L. , Peggs G.N., *The pressure balance - Theory and practice,* National Physical Laboratory, London: Her Majesty's stationery Office, 1982.

[5] R. Giovannozzi, *Costruzione di Macchine,* Vol II, Patron- Bologna , Italy, 1990.

[6] G.F. Molinar, F. Pavese, *Modern gas-Based temperature and pressure measurements,* International cryogenics monograph series, General Editors (K.D. Timmerhaus, Alan F.Clark, Carlo Rizzuto), 1992.

[7] W. Sabuga, et al., *Comparison of methods for calculating distortion in pressure balances up to 400 Mpa EUROMET Project # 256* – January 1998.

[8] W. Sabuga, et al., *Calculation of the Distortion Coefficient and Associated Uncertainty of a PTB 1 GPa Pressure Balance Using Finite Element Analysis* –EUROMET Project 463, Proc. IMEKO TC16 Pressure Symposium Beijing, 2003 pgg. 92-104.

[9] W. Sabuga, et al., *Calculation of the distortion coefficient and associated uncertainty of PTB and LNE 1GPa pressure balances using finite element analysis*—EUROMET project 463, Metrologia 42 (2005) S202–S206.

[10] G. Buonanno et al., *The Piston Distortion Coefficient of Clearance Controlled Pressure Balances Evaluated by Using An Analytical Method* IMEKO Beijing, 2003.